\documentclass[twocolumn,english]{revtex4-1}
\usepackage[T1]{fontenc}
\usepackage[latin9]{inputenc}
\usepackage{amsmath}
\usepackage{amssymb}
\usepackage{graphicx}

\makeatletter
\usepackage{babel}

\usepackage{babel}

\makeatother

\usepackage{babel}
\begin{document}

\title{Quantum Walks as simulators of neutrino oscillations in vacuum and
matter}

\author{G. Di Molfetta$^{1,2}$}

\email{giuseppe.dimolfetta@lif.univ-mrs.fr}

\author{A. Pérez$^{1}$}

\affiliation{$^{1}$Departamento de Física Teórica and IFIC, Universidad de Valencia-CSIC,
Dr. Moliner 50, 46100-Burjassot, Spain\\
$^{2}$Aix-Marseille Université, CNRS, LIF, Marseille, France}
\begin{abstract}
We analyze the simulation of Dirac neutrino oscillations using quantum
walks, both in vacuum and in matter. We show that this simulation,
in the continuum limit, reproduces a set of coupled Dirac equations
that describe neutrino flavor oscillations, and we make use of this
to establish a connection with neutrino phenomenology, thus allowing
one to fix the parameters of the simulation for a given neutrino experiment.
We also analyze how matter effects for neutrino propagation can be
simulated in the quantum walk. In this way, important features, such
as the MSW effect, can be incorporated. Thus, the simulation of neutrino
oscillations with the help of quantum walks might be useful to illustrate
these effects in extreme conditions, such as the solar interior or
supernovae.
\end{abstract}
\maketitle

\section{Introduction}

Quantum simulation is important for the ability to explore the behavior
of a quantum system under conditions that are experimentally difficult
to access. A clear example is the simulation of the Dirac equation,
which allows one for the visualization of effects like the zitterbewegung
or the Klein paradox \cite{PhysRevLett.98.253005,Gerritsma2010,DMD14},
which can in fact be easily simulated on a classical computer, but
are hard to verify on the laboratory. Related to this application
of quantum simulations, we will consider in this paper another problem,
that concerns the simulation of neutrino oscillations using the quantum
walk (QW). Neutrino oscillations were proposed by B. Pontecorvo in
1957 \cite{Pontecorvo1957}, under the form of neutrino-antineutrino
oscillations, in analogy to Kaon oscillations, and afterwards as \textquotedbl{}flavor
transitions'' by Maki, Nakagawa and Sakata in 1962 \cite{Maki01111962}.
This kind of oscillations appear because the three types of flavor
states for neutrinos ($\nu_{e}$, $\nu_{\mu}$ and $\nu_{\tau}$)
that have a definite interaction in the Standard Model of particle
interactions (SM) do not coincide with the mass eigenstates of the
Hamiltonian described by the same model. Instead, there is a unitary
transformation that relates both sets of states. As a consequence,
a neutrino which is produced e.g. as an electron neutrino $\nu_{e}$
can be detected, with a given probability, as any of the three flavors
at a later time (see, for example \cite{RevModPhys.59.671,Balantekin2013a}).
Neutrino oscillations have given rise to very rich phenomena, ranging
from the solution to the Solar neutrino problem, supernovae \cite{Raffelt:1996wa},
reactor neutrinos \cite{Kim:2013vda}, the Early Universe \cite{Lesgourgues-Mangano-Miele-Pastor-2013},
or atmospheric neutrinos \cite{Kajita:2004ga}.

QWs are quantum cellular automata in the one particle sector that
can be viewed as formal generalizations of classical random walks.
They have been first considered by Feynman \cite{FeynHibbs65a} as
a possible discretization of the free Dirac dynamics in flat space-time,
and later introduced in the physics literature in a systematic way
by Meyer \cite{Meyer96a}, following the seminal work of Aharonov
\cite{Aharonov93}. A continuous-time version first appeared in \cite{PhysRevLett.102.180501}.
QWs have been realized experimentally with a wide range of physical
objects and setups \cite{Schmitz09a,PhysRevLett.104.100503,PhysRevLett.104.050502,Karski2009,PhysRevLett.108.010502,PhysRevA.67.042305,PhysRevLett.100.170506},
and are studied in a large variety of contexts, ranging from fundamental
quantum physics \cite{PhysRevLett.100.170506,var96a} to quantum algorithmics
\cite{Amb07a,MNRS07a}, solid-state physics \cite{Aslangul05a,Bose03a,Burg06a,Bose07a}
and biophysics \cite{Collini10a,Engel07a}. Following the Feynman's
idea, several authors have studied the connection between quantum
automata and quantum field theory \cite{PhysRevD.49.6920,Bisio2015}.
In particular, it is well known that the continuous limit of various
QWs formally coincides with the Dirac equation \cite{Strauch06b,Strauch06a,Chandrashekar2013}.
It has been shown recently that several QWs can model the dynamics
of free Dirac fermions coupled to electromagnetic \cite{DMD12a,ADmag16}
and relativistic gravitational fields \cite{DMD13b,DMD14,arrighi2015quantum,Succi2015}. 

Here, we will make use of the capabilities of the QW as discretizations
of quantum field theories, to establish a connection between QWs and
the phenomenology of neutrino oscillations in different scenarios.
Given the known bounds to the neutrino masses, we will concentrate
in relativistic neutrinos, an approximation that holds in the vast
majority of experimental situations. Neutrino oscillations have been
demonstrated by an optical analogical experiment based on the two-state
system of polarized photons traveling through a birefringent crystal
\cite{Weinheimer2010}. In \cite{Noh2011} the authors discuss the
idea of simulating the coupled Dirac equations describing the neutrino
propagation with trapped ions. A different proposal \cite{PhysRevLett.113.150401}
consists in using waveguide arrays. The QW, however, offers the possibility
to perform the simulation by using a suitable modification of some
of the experimental setups already available, or proposed, for the
QW dynamics. Experimental procedures allowing the implementation of
DTQWs with wave functions having more than two components have been
proposed in \cite{PhysRevLett.107.253001,PhysRevA.92.040302}. In
these procedures, the DTQWs are implemented with single photons or
classical light, for example in optical cavities or with atoms in
optical lattices.

A previous work has already motivated such study \cite{Mallick2016}.
Here, we extend this work in two crucial directions. First, we analyze
the continuous version (in space and time) of the proposed model.
In this way, we can obtain the resulting equations that describe the
well-known Hamiltonian dynamics of (relativistic) neutrinos. This
result, not only is a verification of the model as describing neutrino
oscillations, but also allows one to establish a clear correspondence
between the parameters of the simulation and the neutrino data for
a given experiment. Secondly, we show how matter effects can be incorporated
into the simulation. As it is well known, matter effects are crucial
for the explanation of the Solar neutrino problem, specially for the
higher energy $^{8}B$ neutrinos \cite{Balantekin2013a}, and for
long baseline experiments \cite{Bernabeu2002,Feldman:2013vca}. A
quantum simulator can then be used as an extra tool to visualize these
effects from the point of view of a discretized field theory.

This paper is organized as follows. In Sec. II we give a short introduction
to the phenomenon of neutrino oscillations, and illustrate it for
the case of two flavors. In Sec. III we define the dynamics of the
QW model, and we show that, in the continuous space-time limit, we
recover the dynamics of a set of coupled Dirac equations that can
be put in correspondence with the Hamiltonian formulation of neutrino
oscillations, thus allowing one to identify the value of the parameters
needed to simulate a given neutrino experiment. We make use of this
correspondence, in Sec. IV, to give an example of the simulation of
three flavors in vacuum. In Sec. V, we show how matter effects in
the neutrino propagation can be incorporated in the QW model. Our
main conclusions are summarized in Sec. VI. Along this paper we use
natural units, defined by $\hbar=c=1$.

\section{Neutrino oscillations}

Neutrinos are produced as ``flavor states'' via charged or neutral
currents in nuclear reactors, stars, cosmic rays and many other scenarios,
in the way described by the SM. These flavor states $\tilde{\Psi}_{\alpha}$
($\alpha=e,\mu,\tau$) are related, at a given space-point $(t,x)$,
to the mass eigenstates by an unitary transformation $R$, such that
\begin{equation}
\tilde{\Psi}_{\alpha}(t,x)=\sum_{i}R_{\alpha i}\,\Psi_{i}(t,x),\label{3numix}
\end{equation}
where $\Psi_{i}(t,x)$ is the neutrino field with mass $m_{i}$. The
field can, in principle, take either a Dirac or a Majorana form although
we will consider, for definiteness, a Dirac field: The precise nature
of the neutrino field is still undetermined and, although important
for the fundamental knowledge of neutrinos, is irrelevant in the case
of relativistic neutrinos considered here. Eq. (\ref{3numix}) actually
refers to the left chiral components of the neutrinos, since only
these components interact within the SM. We will restrict ourselves
to propagation along one spatial dimension, then $x$ refers to the
coordinate along that dimension.

We will concentrate on neutrino eigenstates, with momentum $\vec{k}$
and energy $E_{i}=\sqrt{\vec{k}^{2}+m_{i}^{2}}$ of the Hamiltonian
$H$ of the theory: 
\begin{equation}
H~|\nu_{i}\rangle=E_{i}~|\nu_{i}\rangle.\label{eigenstate}
\end{equation}
Suppose that, at time $t=0$, a flavor neutrino $|\nu_{\alpha}\rangle$
is produced, then at time $t$ the neutrino state evolves according
to 
\begin{equation}
|\nu_{\alpha}\rangle_{t}=e^{-iHt}\sum_{i=1}^{3}R_{\alpha i}^{*}~|\nu_{i}\rangle=\sum_{i}|\nu_{i}\rangle e^{-iE_{i}t}~R_{\alpha i}^{*}.\label{Schrod1}
\end{equation}
Therefore, at time $t$, the initial neutrino can be detected as any
flavor $\nu_{\beta}$. The probability of the $\nu_{\alpha}\to\nu_{\beta}$
transition after a time $t$ is then given by the following expression
\begin{equation}
P(\nu_{\alpha}\to\nu_{\beta};t)=|\sum_{i=1}^{3}R_{\beta i}~e^{-iE_{i}~t}~R_{\alpha i}^{*}|^{2}.\label{Schrod3}
\end{equation}
To illustrate the latter formula, let us analyze the case with only
two neutrino states (some interesting cases, such as solar and atmospheric
neutrinos, can be effectively described in this way). The matrix $R$
can be taken as 
\begin{eqnarray}
R=\left(\begin{array}{cc}
\cos\phi_{12} & \sin\phi_{12}\\
-\sin\phi_{12} & \cos\phi_{12}
\end{array}\right),\label{2nu1}
\end{eqnarray}
$\phi_{12}$ being the mixing angle. By approximating $E_{i}\simeq|\vec{k}|+\frac{m_{i}^{2}}{2|\vec{k}|}$,
Eq. (\ref{Schrod3}) gives, for $\beta\neq\alpha$: 
\begin{equation}
P(\nu_{\alpha}\to\nu_{\beta};t)=\sin^{2}2\phi_{12}\sin^{2}\varphi(t),\label{Probtrans}
\end{equation}
where 
\begin{equation}
\varphi(t)\equiv\frac{\Delta m_{12}^{2}t}{2|\vec{k}|},\label{phaseDirac}
\end{equation}
with the notation $\Delta m_{12}^{2}=m_{2}^{2}-m_{1}^{2}$. Notice
that, in the latter equation, one can replace $|\vec{k}|$ by $E$
within the same order of approximation as above, where $E=E_{1}\simeq E_{2}$.
Also, it is customary to use the distance $L\simeq t$ traveled by
the neutrinos from the production to the detection point. Then, the
argument of the oscillatory function in (\ref{Probtrans}) becomes
\begin{equation}
\varphi(L)\simeq5.08\frac{\Delta m_{12}^{2}L}{2E},\label{oscilphasenumerics}
\end{equation}
when $\Delta m_{12}^{2}$ is expressed in $eV^{2}$, $L$ in Km and
$E$ in GeV units.

\section{The model}

Consider a QW defined over discrete time and discrete one dimensional
space, labeled respectively by $j\in\mathbb{N}$ and $p\in\mathbb{Z}$.
This QW is driven by an homogeneous coin acting on the $2n$-dimensional
tensor-product Hilbert space $\mathcal{H}=\mathcal{H}_{\text{spin}}\otimes\mathcal{H}_{n}$,
where $\mathcal{H}_{\text{spin}}$ is $2$-dimensional and $\mathcal{H}_{n}$
describes the n-flavor Hilbert space of the walker. The evolution
equations read 
\begin{equation}
\begin{bmatrix}\psi_{j+1,p}^{1}\\
...\\
\psi_{j+1,p}^{n}
\end{bmatrix}\ =\left(\bigoplus_{h=1,n}SQ_{\epsilon}^{h}\right)\begin{bmatrix}\psi_{j,p}^{1}\\
...\\
\psi_{j,p}^{n}
\end{bmatrix},\label{eq:defwalkdiscr}
\end{equation}
where $Q_{\epsilon}^{h}$ $\in SU(2)$, $h=1...n$, and 
\begin{equation}
Q_{\epsilon}^{h}=\begin{pmatrix}\cos(\epsilon\theta_{h}) & i\sin(\epsilon\theta_{h})\\
i\sin(\epsilon\theta_{h}) & \cos(\epsilon\theta_{h}),
\end{pmatrix}
\end{equation}
is the quantum coin acting on each flavor state of the walker depending
on dimensionless real parameters $(\epsilon,\theta_{h})$. The operator
$S$ is the usual spin-dependent translation acting on each flavor
component, $\psi_{j,p}^{h}=\{\psi_{j,p}^{\uparrow h},\psi_{j,p}^{\downarrow h}\}\in\mathcal{H}_{\text{spin}}$
and defined as follows: 
\begin{equation}
S\psi_{j,p}^{h}=\left(\psi_{j,p+1}^{\uparrow h},\psi_{j,p-1}^{\downarrow h}\right)^{\top}.
\end{equation}
We note $\Psi=(\bigoplus_{h=1,n}\psi^{h})^{\top}\in\mathcal{H}_{n}$.
\\
 Equations (\ref{eq:defwalkdiscr}) describe the evolution of $n$
independent two-level systems, and it has been shown that each of
them recover, in the continuous limit, the Dirac equation \cite{DMD12a,DMD13b,DMD14},
where the parameter $\theta_{h}$ corresponds to the mass of the fermion.
Let us now consider an additional unitary operator $R$ and its inverse
$R^{-1}=R^{\dagger}$ acting on $\mathcal{H}_{n}\otimes\mathcal{H}_{2}$.
The specific role of this operator is to mix the flavor degrees of
freedom, as in Eq. (\ref{3numix}). The evolution equation becomes:
\begin{equation}
\tilde{\Psi}_{j+1,p}=R\left(\bigoplus_{h=1,n}SQ_{\epsilon}^{h}\right)R^{\dagger}\tilde{\Psi}_{j,p}\label{eq:walkwithtilde}
\end{equation}
where we call $\tilde{\Psi}_{j,p}=R\Psi_{j,p}$ the flavor eigenstates.
In the next section we will prove that this simple model recovers,
in the continuous limit, the oscillatory dynamical behavior of Dirac
neutrinos in vacuum.

\subsection{Generalized Dirac equation for neutrinos}

In order to compute the continuous limit of equations (\ref{eq:defwalkdiscr}),
we first consider that $\tilde{\Psi}_{j,p}$ are the `values' $\tilde{\Psi}(t_{j},x_{p})$
taken by a two-component wave-function $\tilde{\Psi}$ at space-time
point $(t_{j}=j\epsilon,x_{p}=p\epsilon)$. We assume that $\tilde{\Psi}$
is at least twice differentiable with respect to both space and time
variables for all sufficiently small values of $\epsilon$. Assuming
the existence of the continuous limit imposes the following constraint
on the coin: 
\begin{equation}
\lim_{\epsilon\rightarrow0}\left[R\left(\bigoplus_{h=1,n}SQ_{\epsilon}^{h}\right)R^{\dagger}\right]=I_{2n}.\label{eq:condition}
\end{equation}
The above equation (\ref{eq:condition}) is directly verified because
$Q^{i}=I_{2}$ as $\epsilon\rightarrow0$ and $RR^{\dagger}$= $I_{2n}$
by definition.\\
 When we Taylor-expand (\ref{eq:walkwithtilde}) at first order in
$\epsilon$, the zero-order terms cancel each other, since (\ref{eq:condition})
is satisfied, and the first-order terms provide the following system
of partial differential equations (PDEs): 
\begin{equation}
\left[\partial_{t}-\left(\bigoplus_{h=1,n}\sigma_{z}\right)\partial_{x}-i\mathcal{M}\right]\tilde{\Psi}(t,x)=0,\label{eq:diracN}
\end{equation}
where $\mathcal{M}_{r,s}$ is the mass tensor, with indices $r,s=\{1,...,n\}$:
\begin{equation}
\mathcal{M}_{r,s}=\sum_{c=1,...,n}R_{r,c}\bar{Q}_{c}R_{r,c}^{-1}\hspace{0.5cm}\bar{Q}_{c}=\bar{\theta}_{c}\sigma_{x}.
\end{equation}
Equation (\ref{eq:diracN}) are standard Dirac equations for $n$
relativistic flavor neutrinos in (1+1) continuous space-time. Notice
that, because $\tilde{\Psi}(t,x)$ is a solution of the Dirac equation,
it is also a solution of the Klein-Gordon (KG) equation. As usual,
we can expand the neutrino field in plane waves of the form $\tilde{\Psi}(t,x)=\tilde{\Psi}_{k}(t)e^{ikx}$,
so that equation (\ref{eq:diracN}) transcribes in $(\partial_{t}^{2}+k^{2}+\mathcal{M}^{\dagger}\mathcal{M})\tilde{\Psi}_{k}(t)=0$,
where $\mathcal{M}^{\dagger}\mathcal{M}$ = $\text{diag}\{\bar{\theta}_{1}^{2},...,\bar{\theta}_{n}^{2}\}$.
To describe relativistic neutrinos, however, the condition $k>>\bar{\theta}_{i}$
has to be fulfilled. In this limit, one can linearize the previous
KG equation as follows: 
\begin{equation}
i\partial_{t}\tilde{\Psi}_{k}(t)-\Omega_{k}\tilde{\Psi}_{k}(t)=0,\hspace{0.2cm}\Omega_{k}=(k+R^{\dagger}\frac{|\mathcal{M}|^{2}}{2k}R),\label{eq:diracLIN}
\end{equation}
where we have considered only the positive-energy probability amplitudes
(neutrinos), and defined $|\mathcal{M}|=\sqrt{\mathcal{M}^{\dagger}\mathcal{M}}$.
For negative-energy states (antineutrinos) a global minus sign appears
in equation (\ref{eq:diracLIN}). Following a similar procedure as
in Sec. II one can obtain, starting from Eq. (\ref{eq:diracLIN}),
the transition probability, which can be written as 
\begin{eqnarray}
P(\nu_{\alpha}\rightarrow\nu_{\beta};t)=|\sum_{k}\tilde{\psi}_{k}^{\alpha*}(0)\tilde{\psi}_{k}^{\beta}(t)|^{2}.
\end{eqnarray}
For example, in the case of two neutrino flavors, one would arrive
to Eq. (\ref{Probtrans}), but with the phase $\varphi(t)$ given
by 
\begin{equation}
\varphi(t)=\frac{\bar{\theta}_{2}^{2}-\bar{\theta}_{1}^{2}}{2k}t.\label{phaseQWcontin}
\end{equation}

\section{Simulation of neutrino oscillations in vacuum}

The above equations were derived for any number of neutrino generations.
Writing them explicitly in terms of mixing angles is extremely cumbersome
in general. Let us consider the three neutrino generations $\{e,\mu,\tau\}$
in vacuum, respectively the electron, the muon and the tau neutrino.
The transformation $R$ recovers the Pontecorvo-Maki-Nakagawa-Sakata
(PMNS) mixing matrix and depends on three real parameters \footnote{We do not include CP symmetry violation effects in this model.}:
\begin{equation}
R=e^{i\phi_{\mu\tau}\lambda_{7}}e^{i\phi_{e\tau}\lambda_{5}}e^{i\phi_{e\mu}\lambda_{2}}
\end{equation}
where the $\lambda$ are the Gell-Mann matrices that correspond to
the spin-one matrices of SO(3). Each angle $\phi_{ij}$ corresponds
to the mixing between two neutrino species. Let us remark that when
$R$ is the identity, the solutions of Eq. (\ref{eq:diracLIN}) are
propagating plane waves $\propto e^{-i(Et-kx)}$, where $k^{2}=E^{2}-\bar{\theta}_{h}^{2}$,
h $\in\{e,\mu,\tau\}$.

The comparison between Eqs. (\ref{phaseDirac}) and (\ref{phaseQWcontin})
allows us to establish a criterion to simulate a given neutrino oscillation
experiment, by requiring that the accumulated phase $\varphi(t)$
takes the same value in both cases, where $t$ represents the number
of timesteps in the QW simulation. For example, one can choose the
values $\bar{\theta}_{h}^{2}$ as given the neutrino square masses
$m_{i}^{2}$, expressed in eV$^{2}$, and $k$ given by the energy
in GeV. This will, in turn, fix the number of required timesteps,
for a given distance $L$. A second condition, of course, is that
the mixing angles $\phi_{ij}$ correspond to the observed values (see
\cite{Forero2014} for recent results).

\begin{figure}[h]
\includegraphics[width=1\columnwidth]{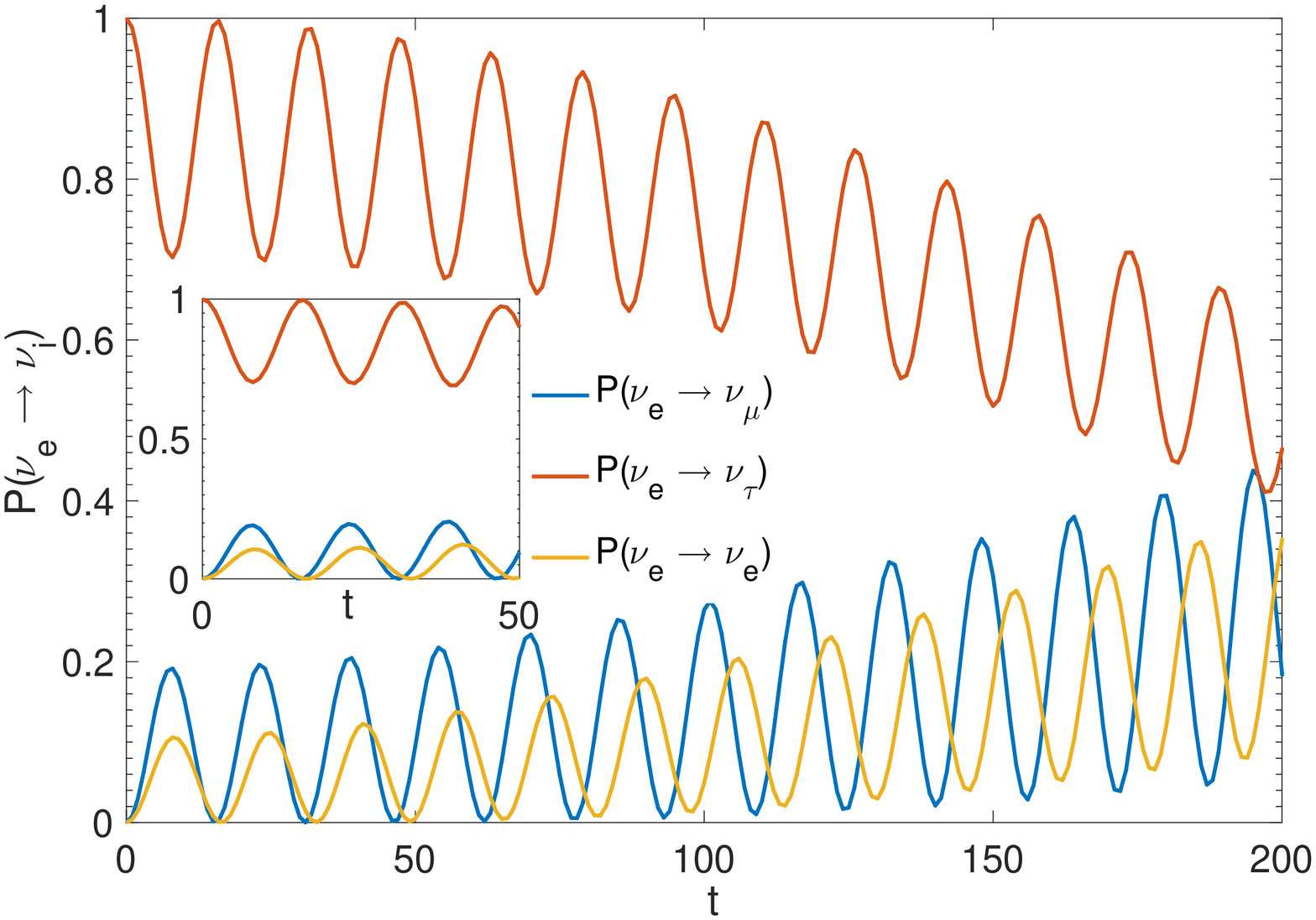} \includegraphics[width=1\columnwidth]{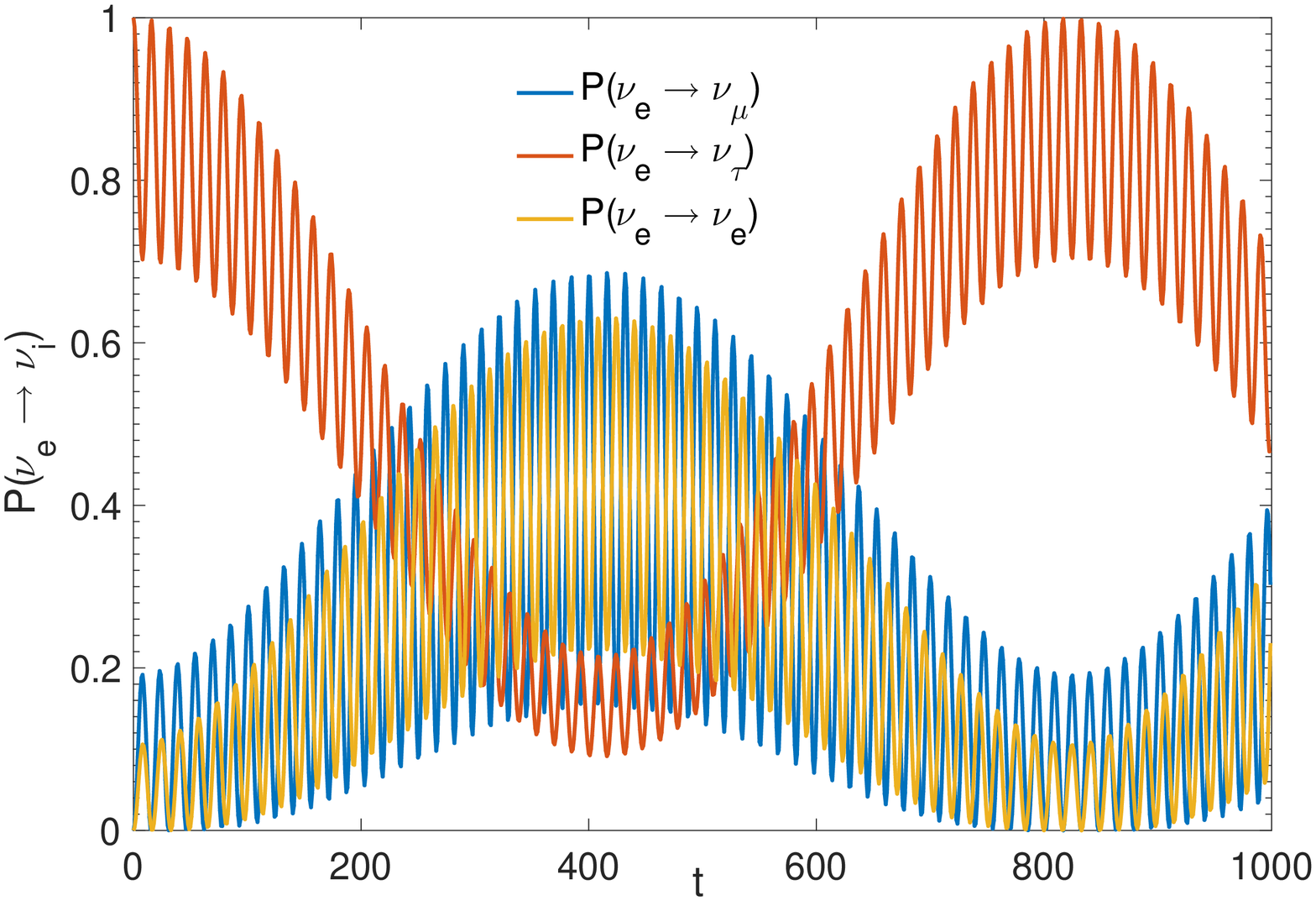}
\caption{(Color online) Time evolution of the probability $P(\nu_{\alpha}\rightarrow\nu_{\beta};t)$,
$\beta\in\{e,\tau,\mu\}$ of a three flavor neutrino oscillation in
vacuum, simulated by a QW. (Top) Short (inset) and medium time range
(200 time steps). (Bottom) Long time range (1000 time steps). The
mass differences are ${\Delta m^{2}}_{e\mu}=0.003$ rad, ${\Delta m^{2}}_{\mu\tau}=0.32$
rad, ${\Delta m^{2}}_{e\tau}=0.31$ rad, and the mixing angles $\phi_{12}=0.34$
rad, $\phi_{13}=0.54$, $\phi_{23}=0.45$ rad, whereas $k_{0}=100$.
The initial condition is defined in Eq. (\ref{eq:CI}). }
\label{fig:traj} 
\end{figure}

In Fig. \ref{fig:traj} we can observe the oscillatory behavior of
three flavor neutrinos starting from a pure electron-neutrino initial
state: 
\begin{equation}
\tilde{\psi}_{k}^{i*}(0)=\frac{1}{\sqrt{n}}\sum_{p=0}^{n-1}e^{-i(k-k_{0})x_{p}}\otimes(1,0,0,0,0,0)^{\top}.\label{eq:CI}
\end{equation}
One can observe that these plots reproduce the ones corresponding
to actual calculations of neutrino oscillations (see for example \cite{invisibleseu}).

\section{Neutrino oscillations in matter}

Wolfenstein (1978) was first to recognize that the medium induces
a modification of the neutrino dispersion relation \cite{Wolfenstein:1978ue-pot},
and then Mikheyev and Smirnov showed that oscillations can be resonant
when neutrinos pass through a density gradient, so that the flavor
branches of the dispersion relations have an avoided crossing \cite{Mikheev1985}.
This Mikheyev-Smirnov-Wolfenstein (MSW) effect is very important in
astrophysics because neutrinos are naturally produced in the interior
of stars and stream through a density gradient into empty space. \\
 In order to model neutrino oscillations in a medium, let us consider
the case with only two flavors, which we freely denote as $\nu_{e}$
and $\nu_{\mu}$. Notice that, as already mentioned, many realistic
experiments can be effectively reduced to a two-neutrino oscillation.
In that case we consider one mixing angle and a $2$-dimensional matrix
$R$: 
\begin{equation}
R=\begin{pmatrix}\cos\phi & \sin\phi\\
-\sin\phi & \cos\phi
\end{pmatrix}\otimes\mathbb{I}_{2}
\end{equation}

Matter interaction is modeled by introducing a position-dependent
phase: 
\begin{equation}
\Psi_{j+1,p}=V_{p}R\left(\bigoplus_{h=1,2}SQ_{\epsilon}^{h}\right)R^{\dagger}\Psi_{j,p},
\end{equation}
where 
\begin{equation}
V_{p}=\mbox{diag}(e^{i\epsilon\rho_{p}},1)\otimes\mathbb{I}_{2}
\end{equation}

By the same procedure presented in section A we can derive the following
Dirac equations: 
\begin{equation}
i\partial_{t}\tilde{\Psi}(t,x)-\mathcal{H}_{m}\tilde{\Psi}(t,x)=0,\label{eq:Diracmatter}
\end{equation}
\begin{equation}
\mathcal{H}_{m}=i\left(\sigma_{z}\otimes\mathcal{I}\right)\partial_{x}-\mathcal{M}+\mathcal{V}(x)\label{eq:Hamiltonianm}
\end{equation}
where the potential $\mathcal{V}$ reads as follows: 
\begin{equation}
\mathcal{V}(x)=\gamma^{5}\mathbb{I}_{4}\rho(x),
\end{equation}
with $\mathbb{I}_{4}$ the $4$-dimensional identity matrix, and $\gamma^{5}=\frac{1}{2}(1+\sigma_{z})$.
The transition probability from an electron neutrino to the muon neutrino
is given by: 
\begin{eqnarray}
P(\nu_{e}\rightarrow\nu_{\mu};t)=|\sum_{h}U_{eh}e^{-i\omega_{h}t}U_{\mu h}^{*}|^{2}
\end{eqnarray}

where $U$ = $U(x)$ is position-dependent and reads: 
\begin{equation}
U(x)=\begin{pmatrix}\cos\Phi(x) & \sin\Phi(x)\\
-\sin\Phi(x) & \cos\Phi(x)
\end{pmatrix},
\end{equation}
and the mixing angle $\Phi(x)$ is related to the mixing angle in
vacuum $\phi$ via the well-known relation: 
\begin{equation}
\sin^{2}2\Phi(x)=\frac{\sin^{2}2\phi}{A(x)},
\end{equation}
with $A(x)=\left[\cos(2\phi)-\frac{2E\rho(x)}{\Delta^{2}m}\right]{}^{2}+\sin^{2}2\phi$
the resonance factor. Notice that, if matter density is not constant,
it is necessary to take into account the effect of $\partial_{x}U(x)$
in the evolution equation in mass eigenstates. In fact: 
\begin{equation}
U\mathcal{H}_{m}U^{-1}=iU\left(\sigma_{z}\otimes\mathcal{I}\right)\partial_{x}U^{-1}+\text{diag. terms}
\end{equation}
\begin{figure}
\includegraphics[width=1\columnwidth]{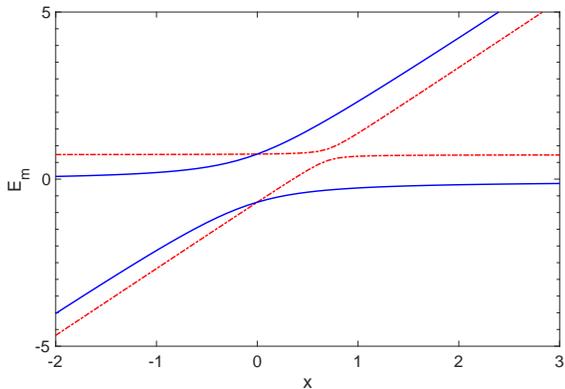}

\caption{Level crossing scheme. Dependence of the eigenvalues of the Hamiltonian
in matter with linear density $\rho(x)=x$ : $E_{1m}$ (lower curve)
and $E_{2m}$ (upper curve) on the position $x$ for two different
values of the mixing angle $\phi=0.12$ (solid blue) and $\phi=0.84$
(dashed red line). The mass difference is ${\Delta m^{2}}_{e\mu}=0.72$
rad ($\theta_{e}=0.1$ rad and $\theta_{\mu}$ = 0.2 rad), and $k_{0}=100$.}

\label{fig:Levelcrossing}
\end{figure}

As mentioned above, the eigenvalues $E_{1m}$ and $E_{2m}$ of the
Hamiltonian (\ref{eq:Hamiltonianm}) present an avoided crossing (see
Fig. \ref{fig:Levelcrossing}) close to tthe resonance. The effect
of the off-diagonal terms proportional to $\partial_{x}\Phi(x)$ in
the evolution is to generate transitions between the energy states
close to the resonance. However, if the derivative is much smaller
with respect to the diagonal terms such transitions are negligible.
To quantify the strength of the off-diagonal terms it is useful to
introduce the adiabaticity parameter: 
\begin{equation}
\gamma=\frac{{\Delta m^{2}}_{e\mu}}{4k|\partial_{x}\Phi(x)|}.
\end{equation}
If $\gamma\gg1$ for all $x$ the evolution is adiabatic, otherwise
the probability of flavor transition reads: 
\begin{equation}
P(\nu_{e}\rightarrow\nu_{\mu};t_{f})=-\frac{1}{2}+(\frac{1}{2}-P_{c})\cos2\Phi(t_{f})\cos2\Phi(t_{i})\label{mattertransprob}
\end{equation}
where $t_{f}$ and $t_{i}$ are the final and initial times (which
can be approximated by the final and initial position $x_{f}$ and
$x_{i}$, for relativistic neutrinos), $P_{c}$ is the crossing probability
for $x_{f}\rightarrow\infty$: 
\begin{equation}
P_{c}=\frac{\exp(-\frac{\pi}{2}\gamma_{r})-\exp(-\frac{\pi}{2}\frac{\gamma_{r}}{\sin\phi^{2}})}{1-\exp(-\frac{\pi}{2}\frac{\gamma_{r}}{\sin\phi^{2}})},
\end{equation}
and $\gamma_{r}$ is the adiabaticity parameter $\gamma$ at the resonance
\cite{Petcov1997}.

\begin{figure}[h]
\includegraphics[width=1\columnwidth]{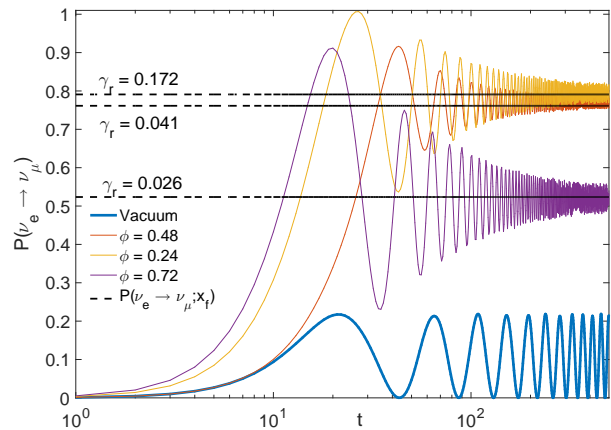} \caption{(Color online) Time evolution of the probability $P(\nu_{\alpha}\rightarrow\nu_{\beta})$
in matter with linear density $\rho(x)=x$ simulated by a QW in 125
time steps. The mass difference and the initial energy is the same
of Fig. \ref{fig:Levelcrossing}. The dashed line (black) represents
the asymptotic crossing probability at the resonance given by formula
(\ref{mattertransprob}), for different adiabaticity parameters $\gamma_{r}$.
The initial and final time steps are $t_{i}=0$, $t_{f}=125$. The
initial condition is as defined in Eq.(\ref{eq:CI}).}
\label{fig:matter} 
\end{figure}

In Fig (\ref{fig:matter}) we have shown that a QW can mimic the time
evolution of two neutrino flavors in matter with a linear density
and for $\gamma_{r}\ll1$. In the long time behavior the transition
probability simulated by the QW converges to the asymptotic probability
$P(\nu_{e}\rightarrow\nu_{\mu};t_{f})$, which confirms the agreement
between the QW's and the neutrino's dynamics in matter.

\section{Conclusions and outlook}

In this paper we have analyzed the simulation of Dirac neutrino oscillations
both in vacuum and in presence of matter effects using quantum walks,
which can therefore be regarded as a discretization of the underlying
field theory. We showed that, in fact, in the continuous limit one
recovers a set of coupled Dirac equations that describe flavor oscillations.
This fact allows us to establish a clear connection with the neutrino
phenomenology for a particular scenario, such as detection of solar,
atmospheric, reactor neutrinos, etc..., so as to fix the relevant
parameters of the simulation (e.g. the initial state, or the necessary
time steps). We have also introduced a way to simulate neutrino propagation
in matter, an element which is crucial for some experiments, such
as detection of solar neutrinos \cite{Balantekin2013a}, or long baseline
experiments \cite{Bernabeu2002,Feldman:2013vca}. As discussed in
Sec. V, the simulation correctly reproduces the flavor conversion
probability for a wide range of values of the adiabaticity parameter.
The above results show that quantum walks can be used to simulate
these phenomena, thus allowing one for a visualization of the neutrino
phenomenology in scenarios like the solar interior or supernovae.

\section{Acknowledgements}

This work has been supported by the Spanish Ministerio de Educación
e Innovación, MICIN-FEDER project FPA2014-54459-P, SEV-2014-0398 and
Generalitat Valenciana grant GVPROMETEOII2014-087. We acknowledge
valuable discussions with E. Roldán and A. Verga.


\begin{thebibliography}{55}%
\makeatletter
\providecommand \@ifxundefined [1]{%
 \@ifx{#1\undefined}
}%
\providecommand \@ifnum [1]{%
 \ifnum #1\expandafter \@firstoftwo
 \else \expandafter \@secondoftwo
 \fi
}%
\providecommand \@ifx [1]{%
 \ifx #1\expandafter \@firstoftwo
 \else \expandafter \@secondoftwo
 \fi
}%
\providecommand \natexlab [1]{#1}%
\providecommand \enquote  [1]{``#1''}%
\providecommand \bibnamefont  [1]{#1}%
\providecommand \bibfnamefont [1]{#1}%
\providecommand \citenamefont [1]{#1}%
\providecommand \href@noop [0]{\@secondoftwo}%
\providecommand \href [0]{\begingroup \@sanitize@url \@href}%
\providecommand \@href[1]{\@@startlink{#1}\@@href}%
\providecommand \@@href[1]{\endgroup#1\@@endlink}%
\providecommand \@sanitize@url [0]{\catcode `\\12\catcode `\$12\catcode
  `\&12\catcode `\#12\catcode `\^12\catcode `\_12\catcode `\%12\relax}%
\providecommand \@@startlink[1]{}%
\providecommand \@@endlink[0]{}%
\providecommand \url  [0]{\begingroup\@sanitize@url \@url }%
\providecommand \@url [1]{\endgroup\@href {#1}{\urlprefix }}%
\providecommand \urlprefix  [0]{URL }%
\providecommand \Eprint [0]{\href }%
\providecommand \doibase [0]{http://dx.doi.org/}%
\providecommand \selectlanguage [0]{\@gobble}%
\providecommand \bibinfo  [0]{\@secondoftwo}%
\providecommand \bibfield  [0]{\@secondoftwo}%
\providecommand \translation [1]{[#1]}%
\providecommand \BibitemOpen [0]{}%
\providecommand \bibitemStop [0]{}%
\providecommand \bibitemNoStop [0]{.\EOS\space}%
\providecommand \EOS [0]{\spacefactor3000\relax}%
\providecommand \BibitemShut  [1]{\csname bibitem#1\endcsname}%
\let\auto@bib@innerbib\@empty
\bibitem [{\citenamefont {Lamata}\ \emph {et~al.}(2007)\citenamefont {Lamata},
  \citenamefont {Le\'on}, \citenamefont {Sch\"atz},\ and\ \citenamefont
  {Solano}}]{PhysRevLett.98.253005}%
  \BibitemOpen
  \bibfield  {author} {\bibinfo {author} {\bibfnamefont {L.}~\bibnamefont
  {Lamata}}, \bibinfo {author} {\bibfnamefont {J.}~\bibnamefont {Le\'on}},
  \bibinfo {author} {\bibfnamefont {T.}~\bibnamefont {Sch\"atz}}, \ and\
  \bibinfo {author} {\bibfnamefont {E.}~\bibnamefont {Solano}},\ }\href
  {\doibase 10.1103/PhysRevLett.98.253005} {\bibfield  {journal} {\bibinfo
  {journal} {Phys. Rev. Lett.}\ }\textbf {\bibinfo {volume} {98}},\ \bibinfo
  {pages} {253005} (\bibinfo {year} {2007})}\BibitemShut {NoStop}%
\bibitem [{\citenamefont {Gerritsma}\ \emph {et~al.}(2010)\citenamefont
  {Gerritsma}, \citenamefont {Kirchmair}, \citenamefont {Zahringer},
  \citenamefont {Solano}, \citenamefont {Blatt},\ and\ \citenamefont
  {Roos}}]{Gerritsma2010}%
  \BibitemOpen
  \bibfield  {author} {\bibinfo {author} {\bibfnamefont {R.}~\bibnamefont
  {Gerritsma}}, \bibinfo {author} {\bibfnamefont {G.}~\bibnamefont
  {Kirchmair}}, \bibinfo {author} {\bibfnamefont {F.}~\bibnamefont
  {Zahringer}}, \bibinfo {author} {\bibfnamefont {E.}~\bibnamefont {Solano}},
  \bibinfo {author} {\bibfnamefont {R.}~\bibnamefont {Blatt}}, \ and\ \bibinfo
  {author} {\bibfnamefont {C.~F.}\ \bibnamefont {Roos}},\ }\href
  {http://dx.doi.org/10.1038/nature08688} {\bibfield  {journal} {\bibinfo
  {journal} {Nature}\ }\textbf {\bibinfo {volume} {463}},\ \bibinfo {pages}
  {68} (\bibinfo {year} {2010})}\BibitemShut {NoStop}%
\bibitem [{\citenamefont {Molfetta}\ \emph {et~al.}(2014)\citenamefont
  {Molfetta}, \citenamefont {Debbasch},\ and\ \citenamefont {Brachet}}]{DMD14}%
  \BibitemOpen
  \bibfield  {author} {\bibinfo {author} {\bibfnamefont {G.~D.}\ \bibnamefont
  {Molfetta}}, \bibinfo {author} {\bibfnamefont {F.}~\bibnamefont {Debbasch}},
  \ and\ \bibinfo {author} {\bibfnamefont {M.}~\bibnamefont {Brachet}},\
  }\href@noop {} {\bibfield  {journal} {\bibinfo  {journal} {Phys. A}\ }\textbf
  {\bibinfo {volume} {397}} (\bibinfo {year} {2014})}\BibitemShut {NoStop}%
\bibitem [{\citenamefont {Pontecorvo}(1957)}]{Pontecorvo1957}%
  \BibitemOpen
  \bibfield  {author} {\bibinfo {author} {\bibfnamefont {B.}~\bibnamefont
  {Pontecorvo}},\ }\href@noop {} {\bibfield  {journal} {\bibinfo  {journal}
  {Sov. Phys. JETP. Zh. Eksp. Teor. Fiz.}\ }\textbf {\bibinfo {volume} {33}},\
  (\bibinfo {year} {1957})}\BibitemShut {NoStop}%
\bibitem [{\citenamefont {Maki}\ \emph {et~al.}(1962)\citenamefont {Maki},
  \citenamefont {Nakagawa},\ and\ \citenamefont {Sakata}}]{Maki01111962}%
  \BibitemOpen
  \bibfield  {author} {\bibinfo {author} {\bibfnamefont {Z.}~\bibnamefont
  {Maki}}, \bibinfo {author} {\bibfnamefont {M.}~\bibnamefont {Nakagawa}}, \
  and\ \bibinfo {author} {\bibfnamefont {S.}~\bibnamefont {Sakata}},\ }\href
  {\doibase 10.1143/PTP.28.870} {\bibfield  {journal} {\bibinfo  {journal}
  {Progress of Theoretical Physics}\ }\textbf {\bibinfo {volume} {28}},\
  \bibinfo {pages} {870} (\bibinfo {year} {1962})}\BibitemShut {NoStop}%
\bibitem [{\citenamefont {Bilenky}\ and\ \citenamefont
  {Petcov}(1987)}]{RevModPhys.59.671}%
  \BibitemOpen
  \bibfield  {author} {\bibinfo {author} {\bibfnamefont {S.~M.}\ \bibnamefont
  {Bilenky}}\ and\ \bibinfo {author} {\bibfnamefont {S.~T.}\ \bibnamefont
  {Petcov}},\ }\href {\doibase 10.1103/RevModPhys.59.671} {\bibfield  {journal}
  {\bibinfo  {journal} {Rev. Mod. Phys.}\ }\textbf {\bibinfo {volume} {59}},\
  \bibinfo {pages} {671} (\bibinfo {year} {1987})}\BibitemShut {NoStop}%
\bibitem [{\citenamefont {Balantekin}\ and\ \citenamefont
  {Haxton}(2013)}]{Balantekin2013a}%
  \BibitemOpen
  \bibfield  {author} {\bibinfo {author} {\bibfnamefont {A.}~\bibnamefont
  {Balantekin}}\ and\ \bibinfo {author} {\bibfnamefont {W.}~\bibnamefont
  {Haxton}},\ }\href {\doibase http://dx.doi.org/10.1016/j.ppnp.2013.03.007}
  {\bibfield  {journal} {\bibinfo  {journal} {Progress in Particle and Nuclear
  Physics}\ }\textbf {\bibinfo {volume} {71}},\ \bibinfo {pages} {150 }
  (\bibinfo {year} {2013})},\ \bibinfo {note} {fundamental Symmetries in the
  Era of the \{LHC\}}\BibitemShut {NoStop}%
\bibitem [{\citenamefont {Raffelt}(1996)}]{Raffelt:1996wa}%
  \BibitemOpen
  \bibfield  {author} {\bibinfo {author} {\bibfnamefont {G.}~\bibnamefont
  {Raffelt}},\ }\href@noop {} {\emph {\bibinfo {title} {Stars as laboratories
  for fundamental physics: The astrophysics of neutrinos, axions, and other
  weakly interacting particles}}}\ (\bibinfo  {publisher} {University of
  Chicago Press},\ \bibinfo {year} {1996})\ pp.\ \bibinfo {pages} {1--664},\
  \bibinfo {note} {iSBN 0-226-70272-3}\BibitemShut {NoStop}%
\bibitem [{\citenamefont {Kim}\ \emph {et~al.}(2013)\citenamefont {Kim},
  \citenamefont {Lasserre},\ and\ \citenamefont {Wang}}]{Kim:2013vda}%
  \BibitemOpen
  \bibfield  {author} {\bibinfo {author} {\bibfnamefont {S.-B.}\ \bibnamefont
  {Kim}}, \bibinfo {author} {\bibfnamefont {T.}~\bibnamefont {Lasserre}}, \
  and\ \bibinfo {author} {\bibfnamefont {Y.}~\bibnamefont {Wang}},\ }\href
  {\doibase 10.1155/2013/453816} {\bibfield  {journal} {\bibinfo  {journal}
  {Adv.High Energy Phys.}\ }\textbf {\bibinfo {volume} {2013}},\ \bibinfo
  {pages} {453816} (\bibinfo {year} {2013})}\BibitemShut {NoStop}%
\bibitem [{\citenamefont {Lesgourgues}\ \emph {et~al.}(2013)\citenamefont
  {Lesgourgues}, \citenamefont {Mangano}, \citenamefont {Miele},\ and\
  \citenamefont {Pastor}}]{Lesgourgues-Mangano-Miele-Pastor-2013}%
  \BibitemOpen
  \bibfield  {author} {\bibinfo {author} {\bibfnamefont {J.}~\bibnamefont
  {Lesgourgues}}, \bibinfo {author} {\bibfnamefont {G.}~\bibnamefont
  {Mangano}}, \bibinfo {author} {\bibfnamefont {G.}~\bibnamefont {Miele}}, \
  and\ \bibinfo {author} {\bibfnamefont {S.}~\bibnamefont {Pastor}},\ }\href
  {\doibase 10.1017/CBO9781139012874} {\emph {\bibinfo {title} {Neutrino
  Cosmology}}}\ (\bibinfo  {publisher} {Cambridge University Press},\ \bibinfo
  {year} {2013})\ \bibinfo {note} {iSBN 9781139012874}\BibitemShut {NoStop}%
\bibitem [{\citenamefont {Kajita}(2004)}]{Kajita:2004ga}%
  \BibitemOpen
  \bibfield  {author} {\bibinfo {author} {\bibfnamefont {T.}~\bibnamefont
  {Kajita}},\ }\href@noop {} {\bibfield  {journal} {\bibinfo  {journal} {New J.
  Phys.}\ }\textbf {\bibinfo {volume} {6}},\ \bibinfo {pages} {194} (\bibinfo
  {year} {2004})}\BibitemShut {NoStop}%
\bibitem [{\citenamefont {Feynman}\ and\ \citenamefont
  {Hibbs}(1965)}]{FeynHibbs65a}%
  \BibitemOpen
  \bibfield  {author} {\bibinfo {author} {\bibfnamefont {R.}~\bibnamefont
  {Feynman}}\ and\ \bibinfo {author} {\bibfnamefont {A.}~\bibnamefont
  {Hibbs}},\ }\href@noop {} {\bibfield  {journal} {\bibinfo  {journal}
  {International Series in Pure and Applied Physics. McGraw-Hill Book Company}\
  } (\bibinfo {year} {1965})}\BibitemShut {NoStop}%
\bibitem [{\citenamefont {Meyer}(1996)}]{Meyer96a}%
  \BibitemOpen
  \bibfield  {author} {\bibinfo {author} {\bibfnamefont {D.}~\bibnamefont
  {Meyer}},\ }\href@noop {} {\bibfield  {journal} {\bibinfo  {journal} {{J}.
  {S}tat. {P}hys.}\ }\textbf {\bibinfo {volume} {85}} (\bibinfo {year}
  {1996})}\BibitemShut {NoStop}%
\bibitem [{\citenamefont {Aharonov}\ \emph {et~al.}(1993)\citenamefont
  {Aharonov}, \citenamefont {Davidovich},\ and\ \citenamefont
  {Zagury}}]{Aharonov93}%
  \BibitemOpen
  \bibfield  {author} {\bibinfo {author} {\bibfnamefont {Y.}~\bibnamefont
  {Aharonov}}, \bibinfo {author} {\bibfnamefont {L.}~\bibnamefont
  {Davidovich}}, \ and\ \bibinfo {author} {\bibfnamefont {N.}~\bibnamefont
  {Zagury}},\ }\href {\doibase 10.1103/PhysRevA.48.1687} {\bibfield  {journal}
  {\bibinfo  {journal} {Phys. Rev. A}\ }\textbf {\bibinfo {volume} {48}},\
  \bibinfo {pages} {1687} (\bibinfo {year} {1993})}\BibitemShut {NoStop}%
\bibitem [{\citenamefont {Childs}(2009)}]{PhysRevLett.102.180501}%
  \BibitemOpen
  \bibfield  {author} {\bibinfo {author} {\bibfnamefont {A.~M.}\ \bibnamefont
  {Childs}},\ }\href {\doibase 10.1103/PhysRevLett.102.180501} {\bibfield
  {journal} {\bibinfo  {journal} {Phys. Rev. Lett.}\ }\textbf {\bibinfo
  {volume} {102}},\ \bibinfo {pages} {180501} (\bibinfo {year}
  {2009})}\BibitemShut {NoStop}%
\bibitem [{\citenamefont {Schmitz}\ \emph {et~al.}(2009)\citenamefont
  {Schmitz}, \citenamefont {Matjeschk}, \citenamefont {Schneider},
  \citenamefont {Glueckert}, \citenamefont {Enderlein}, \citenamefont {Huber},\
  and\ \citenamefont {Schaetz}}]{Schmitz09a}%
  \BibitemOpen
  \bibfield  {author} {\bibinfo {author} {\bibfnamefont {H.}~\bibnamefont
  {Schmitz}}, \bibinfo {author} {\bibfnamefont {R.}~\bibnamefont {Matjeschk}},
  \bibinfo {author} {\bibfnamefont {C.}~\bibnamefont {Schneider}}, \bibinfo
  {author} {\bibfnamefont {J.}~\bibnamefont {Glueckert}}, \bibinfo {author}
  {\bibfnamefont {M.}~\bibnamefont {Enderlein}}, \bibinfo {author}
  {\bibfnamefont {T.}~\bibnamefont {Huber}}, \ and\ \bibinfo {author}
  {\bibfnamefont {T.}~\bibnamefont {Schaetz}},\ }\href@noop {} {\bibfield
  {journal} {\bibinfo  {journal} {Phys. Rev. Lett.}\ }\textbf {\bibinfo
  {volume} {103}},\ \bibinfo {pages} {090504} (\bibinfo {year}
  {2009})}\BibitemShut {NoStop}%
\bibitem [{\citenamefont {Z\"ahringer}\ \emph {et~al.}(2010)\citenamefont
  {Z\"ahringer}, \citenamefont {Kirchmair}, \citenamefont {Gerritsma},
  \citenamefont {Solano}, \citenamefont {Blatt},\ and\ \citenamefont
  {Roos}}]{PhysRevLett.104.100503}%
  \BibitemOpen
  \bibfield  {author} {\bibinfo {author} {\bibfnamefont {F.}~\bibnamefont
  {Z\"ahringer}}, \bibinfo {author} {\bibfnamefont {G.}~\bibnamefont
  {Kirchmair}}, \bibinfo {author} {\bibfnamefont {R.}~\bibnamefont
  {Gerritsma}}, \bibinfo {author} {\bibfnamefont {E.}~\bibnamefont {Solano}},
  \bibinfo {author} {\bibfnamefont {R.}~\bibnamefont {Blatt}}, \ and\ \bibinfo
  {author} {\bibfnamefont {C.}~\bibnamefont {Roos}},\ }\href {\doibase
  10.1103/PhysRevLett.104.100503} {\bibfield  {journal} {\bibinfo  {journal}
  {Phys. Rev. Lett.}\ }\textbf {\bibinfo {volume} {104}},\ \bibinfo {pages}
  {100503} (\bibinfo {year} {2010})}\BibitemShut {NoStop}%
\bibitem [{\citenamefont {Schreiber}\ \emph {et~al.}(2010)\citenamefont
  {Schreiber}, \citenamefont {Cassemiro}, \citenamefont {Potocek},
  \citenamefont {G\'abris}, \citenamefont {Mosley}, \citenamefont {Andersson},
  \citenamefont {Jex},\ and\ \citenamefont
  {Silberhorn}}]{PhysRevLett.104.050502}%
  \BibitemOpen
  \bibfield  {author} {\bibinfo {author} {\bibfnamefont {A.}~\bibnamefont
  {Schreiber}}, \bibinfo {author} {\bibfnamefont {K.}~\bibnamefont
  {Cassemiro}}, \bibinfo {author} {\bibfnamefont {V.}~\bibnamefont {Potocek}},
  \bibinfo {author} {\bibfnamefont {A.}~\bibnamefont {G\'abris}}, \bibinfo
  {author} {\bibfnamefont {P.}~\bibnamefont {Mosley}}, \bibinfo {author}
  {\bibfnamefont {E.}~\bibnamefont {Andersson}}, \bibinfo {author}
  {\bibfnamefont {I.}~\bibnamefont {Jex}}, \ and\ \bibinfo {author}
  {\bibfnamefont {C.}~\bibnamefont {Silberhorn}},\ }\href {\doibase
  10.1103/PhysRevLett.104.050502} {\bibfield  {journal} {\bibinfo  {journal}
  {Phys. Rev. Lett.}\ }\textbf {\bibinfo {volume} {104}},\ \bibinfo {pages}
  {050502} (\bibinfo {year} {2010})}\BibitemShut {NoStop}%
\bibitem [{\citenamefont {Karski}\ \emph {et~al.}(2009)\citenamefont {Karski},
  \citenamefont {Forster}, \citenamefont {Choi}, \citenamefont {Steffen},
  \citenamefont {Alt}, \citenamefont {Meschede},\ and\ \citenamefont
  {Widera}}]{Karski2009}%
  \BibitemOpen
  \bibfield  {author} {\bibinfo {author} {\bibfnamefont {M.}~\bibnamefont
  {Karski}}, \bibinfo {author} {\bibfnamefont {L.}~\bibnamefont {Forster}},
  \bibinfo {author} {\bibfnamefont {J.-M.}\ \bibnamefont {Choi}}, \bibinfo
  {author} {\bibfnamefont {A.}~\bibnamefont {Steffen}}, \bibinfo {author}
  {\bibfnamefont {W.}~\bibnamefont {Alt}}, \bibinfo {author} {\bibfnamefont
  {D.}~\bibnamefont {Meschede}}, \ and\ \bibinfo {author} {\bibfnamefont
  {A.}~\bibnamefont {Widera}},\ }\href {\doibase 10.1126/science.1174436}
  {\bibfield  {journal} {\bibinfo  {journal} {Science}\ }\textbf {\bibinfo
  {volume} {325}},\ \bibinfo {pages} {174} (\bibinfo {year}
  {2009})}\BibitemShut {NoStop}%
\bibitem [{\citenamefont {Sansoni}\ \emph {et~al.}(2012)\citenamefont
  {Sansoni}, \citenamefont {Sciarrino}, \citenamefont {Vallone}, \citenamefont
  {Mataloni}, \citenamefont {Crespi}, \citenamefont {Ramponi},\ and\
  \citenamefont {Osellame}}]{PhysRevLett.108.010502}%
  \BibitemOpen
  \bibfield  {author} {\bibinfo {author} {\bibfnamefont {L.}~\bibnamefont
  {Sansoni}}, \bibinfo {author} {\bibfnamefont {F.}~\bibnamefont {Sciarrino}},
  \bibinfo {author} {\bibfnamefont {G.}~\bibnamefont {Vallone}}, \bibinfo
  {author} {\bibfnamefont {P.}~\bibnamefont {Mataloni}}, \bibinfo {author}
  {\bibfnamefont {A.}~\bibnamefont {Crespi}}, \bibinfo {author} {\bibfnamefont
  {R.}~\bibnamefont {Ramponi}}, \ and\ \bibinfo {author} {\bibfnamefont
  {R.}~\bibnamefont {Osellame}},\ }\href {\doibase
  10.1103/PhysRevLett.108.010502} {\bibfield  {journal} {\bibinfo  {journal}
  {Phys. Rev. Lett.}\ }\textbf {\bibinfo {volume} {108}},\ \bibinfo {pages}
  {010502} (\bibinfo {year} {2012})}\BibitemShut {NoStop}%
\bibitem [{\citenamefont {Sanders}\ \emph {et~al.}(2003)\citenamefont
  {Sanders}, \citenamefont {Bartlett}, \citenamefont {Tregenna},\ and\
  \citenamefont {Knight}}]{PhysRevA.67.042305}%
  \BibitemOpen
  \bibfield  {author} {\bibinfo {author} {\bibfnamefont {B.}~\bibnamefont
  {Sanders}}, \bibinfo {author} {\bibfnamefont {S.}~\bibnamefont {Bartlett}},
  \bibinfo {author} {\bibfnamefont {B.}~\bibnamefont {Tregenna}}, \ and\
  \bibinfo {author} {\bibfnamefont {P.}~\bibnamefont {Knight}},\ }\href
  {\doibase 10.1103/PhysRevA.67.042305} {\bibfield  {journal} {\bibinfo
  {journal} {Phys. Rev. A}\ }\textbf {\bibinfo {volume} {67}},\ \bibinfo
  {pages} {042305} (\bibinfo {year} {2003})}\BibitemShut {NoStop}%
\bibitem [{\citenamefont {Perets}\ \emph {et~al.}(2008)\citenamefont {Perets},
  \citenamefont {Lahini}, \citenamefont {Pozzi}, \citenamefont {Sorel},
  \citenamefont {Morandotti},\ and\ \citenamefont
  {Silberberg}}]{PhysRevLett.100.170506}%
  \BibitemOpen
  \bibfield  {author} {\bibinfo {author} {\bibfnamefont {H.}~\bibnamefont
  {Perets}}, \bibinfo {author} {\bibfnamefont {Y.}~\bibnamefont {Lahini}},
  \bibinfo {author} {\bibfnamefont {F.}~\bibnamefont {Pozzi}}, \bibinfo
  {author} {\bibfnamefont {M.}~\bibnamefont {Sorel}}, \bibinfo {author}
  {\bibfnamefont {R.}~\bibnamefont {Morandotti}}, \ and\ \bibinfo {author}
  {\bibfnamefont {Y.}~\bibnamefont {Silberberg}},\ }\href {\doibase
  10.1103/PhysRevLett.100.170506} {\bibfield  {journal} {\bibinfo  {journal}
  {Phys. Rev. Lett.}\ }\textbf {\bibinfo {volume} {100}},\ \bibinfo {pages}
  {170506} (\bibinfo {year} {2008})}\BibitemShut {NoStop}%
\bibitem [{\citenamefont {Giulini}\ \emph {et~al.}(1996)\citenamefont
  {Giulini}, \citenamefont {Joos}, \citenamefont {Kiefer}, \citenamefont
  {Kupsch}, \citenamefont {Stamatescu},\ and\ \citenamefont {Zeh}}]{var96a}%
  \BibitemOpen
  \bibfield  {author} {\bibinfo {author} {\bibfnamefont {D.}~\bibnamefont
  {Giulini}}, \bibinfo {author} {\bibfnamefont {E.}~\bibnamefont {Joos}},
  \bibinfo {author} {\bibfnamefont {C.}~\bibnamefont {Kiefer}}, \bibinfo
  {author} {\bibfnamefont {J.}~\bibnamefont {Kupsch}}, \bibinfo {author}
  {\bibfnamefont {I.-O.}\ \bibnamefont {Stamatescu}}, \ and\ \bibinfo {author}
  {\bibfnamefont {H.}~\bibnamefont {Zeh}},\ }\href@noop {} {\emph {\bibinfo
  {title} {Decoherence and the appearance of a Classical World in Quantum
  Theory}}}\ (\bibinfo  {publisher} {Springer-Verlag},\ \bibinfo {address}
  {Berlin},\ \bibinfo {year} {1996})\BibitemShut {NoStop}%
\bibitem [{\citenamefont {Ambainis}(2007)}]{Amb07a}%
  \BibitemOpen
  \bibfield  {author} {\bibinfo {author} {\bibfnamefont {A.}~\bibnamefont
  {Ambainis}},\ }\href@noop {} {\bibfield  {journal} {\bibinfo  {journal} {SIAM
  Journal on Computing}\ }\textbf {\bibinfo {volume} {37}},\ \bibinfo {pages}
  {210} (\bibinfo {year} {2007})}\BibitemShut {NoStop}%
\bibitem [{\citenamefont {Magniez}\ \emph {et~al.}( ACM)\citenamefont
  {Magniez}, \citenamefont {A.~Nayak},\ and\ \citenamefont {Santha}}]{MNRS07a}%
  \BibitemOpen
  \bibfield  {author} {\bibinfo {author} {\bibfnamefont {F.}~\bibnamefont
  {Magniez}}, \bibinfo {author} {\bibfnamefont {J.~R.}\ \bibnamefont
  {A.~Nayak}}, \ and\ \bibinfo {author} {\bibfnamefont {M.}~\bibnamefont
  {Santha}},\ }\href@noop {} {\bibfield  {journal} {\bibinfo  {journal} {SIAM
  Journal on Computing - Proceedings of the thirty-ninth annual ACM symposium
  on Theory of computing}\ } (\bibinfo {year} {New {Y}ork, 2007.
  ACM.})}\BibitemShut {NoStop}%
\bibitem [{\citenamefont {Aslangul}(2005)}]{Aslangul05a}%
  \BibitemOpen
  \bibfield  {author} {\bibinfo {author} {\bibfnamefont {C.}~\bibnamefont
  {Aslangul}},\ }\href@noop {} {\bibfield  {journal} {\bibinfo  {journal}
  {Journal of Physics A: Mathematical and Theoretical}\ }\textbf {\bibinfo
  {volume} {38}},\ \bibinfo {pages} {1} (\bibinfo {year} {2005})}\BibitemShut
  {NoStop}%
\bibitem [{\citenamefont {Bose}(2003)}]{Bose03a}%
  \BibitemOpen
  \bibfield  {author} {\bibinfo {author} {\bibfnamefont {S.}~\bibnamefont
  {Bose}},\ }\href@noop {} {\bibfield  {journal} {\bibinfo  {journal} {Phys.
  Rev. Lett.}\ }\textbf {\bibinfo {volume} {91}},\ \bibinfo {pages} {207901}
  (\bibinfo {year} {2003})}\BibitemShut {NoStop}%
\bibitem [{\citenamefont {Burgarth}(2006)}]{Burg06a}%
  \BibitemOpen
  \bibfield  {author} {\bibinfo {author} {\bibfnamefont {D.}~\bibnamefont
  {Burgarth}},\ }\href@noop {} {\bibfield  {journal} {\bibinfo  {journal}
  {University College London}\ }\textbf {\bibinfo {volume} {PhD thesis}}
  (\bibinfo {year} {2006})}\BibitemShut {NoStop}%
\bibitem [{\citenamefont {Bose}(2007)}]{Bose07a}%
  \BibitemOpen
  \bibfield  {author} {\bibinfo {author} {\bibfnamefont {S.}~\bibnamefont
  {Bose}},\ }\href@noop {} {\bibfield  {journal} {\bibinfo  {journal} {Contemp.
  {P}hys.}\ }\textbf {\bibinfo {volume} {48}},\ \bibinfo {pages} {13 }
  (\bibinfo {year} {2007})}\BibitemShut {NoStop}%
\bibitem [{\citenamefont {Collini}\ \emph {et~al.}()\citenamefont {Collini},
  \citenamefont {Wong}, \citenamefont {Wilk}, \citenamefont {Curmi},
  \citenamefont {Brumer},\ and\ \citenamefont {Scholes}}]{Collini10a}%
  \BibitemOpen
  \bibfield  {author} {\bibinfo {author} {\bibfnamefont {E.}~\bibnamefont
  {Collini}}, \bibinfo {author} {\bibfnamefont {C.}~\bibnamefont {Wong}},
  \bibinfo {author} {\bibfnamefont {K.}~\bibnamefont {Wilk}}, \bibinfo {author}
  {\bibfnamefont {P.}~\bibnamefont {Curmi}}, \bibinfo {author} {\bibfnamefont
  {P.}~\bibnamefont {Brumer}}, \ and\ \bibinfo {author} {\bibfnamefont
  {G.}~\bibnamefont {Scholes}},\ }\href@noop {} {\bibinfo  {journal} {Nature}\
  ,\ \bibinfo {pages} {644}}\BibitemShut {NoStop}%
\bibitem [{\citenamefont {Engel}\ \emph {et~al.}()\citenamefont {Engel},
  \citenamefont {Calhoun}, \citenamefont {Read}, \citenamefont {Ahn},
  \citenamefont {Manal}, \citenamefont {Cheng}, \citenamefont {Blankenship},\
  and\ \citenamefont {Fleming}}]{Engel07a}%
  \BibitemOpen
\bibfield  {journal} {  }\bibfield  {author} {\bibinfo {author} {\bibfnamefont
  {G.}~\bibnamefont {Engel}}, \bibinfo {author} {\bibfnamefont
  {T.}~\bibnamefont {Calhoun}}, \bibinfo {author} {\bibfnamefont
  {R.}~\bibnamefont {Read}}, \bibinfo {author} {\bibfnamefont {T.-K.}\
  \bibnamefont {Ahn}}, \bibinfo {author} {\bibfnamefont {T.}~\bibnamefont
  {Manal}}, \bibinfo {author} {\bibfnamefont {Y.-C.}\ \bibnamefont {Cheng}},
  \bibinfo {author} {\bibfnamefont {R.}~\bibnamefont {Blankenship}}, \ and\
  \bibinfo {author} {\bibfnamefont {G.~R.}\ \bibnamefont {Fleming}},\
  }\href@noop {} {\bibinfo  {journal} {Nature}\ ,\ \bibinfo {pages}
  {782}}\BibitemShut {NoStop}%
\bibitem [{\citenamefont {Bialynicki-Birula}(1994)}]{PhysRevD.49.6920}%
  \BibitemOpen
\bibfield  {journal} {  }\bibfield  {author} {\bibinfo {author} {\bibfnamefont
  {I.}~\bibnamefont {Bialynicki-Birula}},\ }\href {\doibase
  10.1103/PhysRevD.49.6920} {\bibfield  {journal} {\bibinfo  {journal} {Phys.
  Rev. D}\ }\textbf {\bibinfo {volume} {49}},\ \bibinfo {pages} {6920}
  (\bibinfo {year} {1994})}\BibitemShut {NoStop}%
\bibitem [{\citenamefont {Bisio}\ \emph {et~al.}(2015)\citenamefont {Bisio},
  \citenamefont {D'Ariano},\ and\ \citenamefont {Tosini}}]{Bisio2015}%
  \BibitemOpen
  \bibfield  {author} {\bibinfo {author} {\bibfnamefont {A.}~\bibnamefont
  {Bisio}}, \bibinfo {author} {\bibfnamefont {G.~M.}\ \bibnamefont {D'Ariano}},
  \ and\ \bibinfo {author} {\bibfnamefont {A.}~\bibnamefont {Tosini}},\ }\href
  {\doibase http://dx.doi.org/10.1016/j.aop.2014.12.016} {\bibfield  {journal}
  {\bibinfo  {journal} {Annals of Physics}\ }\textbf {\bibinfo {volume}
  {354}},\ \bibinfo {pages} {244 } (\bibinfo {year} {2015})}\BibitemShut
  {NoStop}%
\bibitem [{\citenamefont {Strauch}(2006{\natexlab{a}})}]{Strauch06b}%
  \BibitemOpen
  \bibfield  {author} {\bibinfo {author} {\bibfnamefont {F.}~\bibnamefont
  {Strauch}},\ }\href@noop {} {\bibfield  {journal} {\bibinfo  {journal} {Phys.
  {R}ev. A}\ }\textbf {\bibinfo {volume} {74}},\ \bibinfo {pages} {030301}
  (\bibinfo {year} {2006}{\natexlab{a}})}\BibitemShut {NoStop}%
\bibitem [{\citenamefont {Strauch}(2006{\natexlab{b}})}]{Strauch06a}%
  \BibitemOpen
  \bibfield  {author} {\bibinfo {author} {\bibfnamefont {F.}~\bibnamefont
  {Strauch}},\ }\href@noop {} {\bibfield  {journal} {\bibinfo  {journal} {Phys.
  {R}ev. A}\ }\textbf {\bibinfo {volume} {73}},\ \bibinfo {pages} {054302}
  (\bibinfo {year} {2006}{\natexlab{b}})}\BibitemShut {NoStop}%
\bibitem [{\citenamefont {Chandrashekar}(2013)}]{Chandrashekar2013}%
  \BibitemOpen
  \bibfield  {author} {\bibinfo {author} {\bibfnamefont {C.~M.}\ \bibnamefont
  {Chandrashekar}},\ }\href@noop {} {\bibfield  {journal} {\bibinfo  {journal}
  {Sci. Rep.}\ }\textbf {\bibinfo {volume} {3}},\ \bibinfo {pages} {2829}
  (\bibinfo {year} {2013})}\BibitemShut {NoStop}%
\bibitem [{\citenamefont {Molfetta}\ and\ \citenamefont
  {Debbasch}(2012)}]{DMD12a}%
  \BibitemOpen
  \bibfield  {author} {\bibinfo {author} {\bibfnamefont {G.~D.}\ \bibnamefont
  {Molfetta}}\ and\ \bibinfo {author} {\bibfnamefont {F.}~\bibnamefont
  {Debbasch}},\ }\href@noop {} {\bibfield  {journal} {\bibinfo  {journal} {J.
  {M}ath. {P}hys.}\ }\textbf {\bibinfo {volume} {53}},\ \bibinfo {pages}
  {123302} (\bibinfo {year} {2012})}\BibitemShut {NoStop}%
\bibitem [{\citenamefont {Arnault}\ and\ \citenamefont
  {Debbasch}(2016)}]{ADmag16}%
  \BibitemOpen
  \bibfield  {author} {\bibinfo {author} {\bibfnamefont {P.}~\bibnamefont
  {Arnault}}\ and\ \bibinfo {author} {\bibfnamefont {F.}~\bibnamefont
  {Debbasch}},\ }\href {\doibase http://dx.doi.org/10.1016/j.physa.2015.08.011}
  {\bibfield  {journal} {\bibinfo  {journal} {Physica A: Statistical Mechanics
  and its Applications}\ }\textbf {\bibinfo {volume} {443}},\ \bibinfo {pages}
  {179 } (\bibinfo {year} {2016})}\BibitemShut {NoStop}%
\bibitem [{\citenamefont {Molfetta}\ \emph {et~al.}(2013)\citenamefont
  {Molfetta}, \citenamefont {Debbasch},\ and\ \citenamefont
  {Brachet}}]{DMD13b}%
  \BibitemOpen
  \bibfield  {author} {\bibinfo {author} {\bibfnamefont {G.~D.}\ \bibnamefont
  {Molfetta}}, \bibinfo {author} {\bibfnamefont {F.}~\bibnamefont {Debbasch}},
  \ and\ \bibinfo {author} {\bibfnamefont {M.}~\bibnamefont {Brachet}},\
  }\href@noop {} {\bibfield  {journal} {\bibinfo  {journal} {Phys. Rev. A}\
  }\textbf {\bibinfo {volume} {88}} (\bibinfo {year} {2013})}\BibitemShut
  {NoStop}%
\bibitem [{\citenamefont {Arrighi}\ \emph {et~al.}(2015)\citenamefont
  {Arrighi}, \citenamefont {Facchini},\ and\ \citenamefont
  {Forets}}]{arrighi2015quantum}%
  \BibitemOpen
  \bibfield  {author} {\bibinfo {author} {\bibfnamefont {P.}~\bibnamefont
  {Arrighi}}, \bibinfo {author} {\bibfnamefont {S.}~\bibnamefont {Facchini}}, \
  and\ \bibinfo {author} {\bibfnamefont {M.}~\bibnamefont {Forets}},\
  }\href@noop {} {\bibfield  {journal} {\bibinfo  {journal} {Quantum
  Information Processing}\ ,\ \bibinfo {pages} {1}} (\bibinfo {year}
  {2015})}\BibitemShut {NoStop}%
\bibitem [{\citenamefont {Succi}\ \emph {et~al.}(2015)\citenamefont {Succi},
  \citenamefont {Fillion-Gourdeau},\ and\ \citenamefont
  {Palpacelli}}]{Succi2015}%
  \BibitemOpen
  \bibfield  {author} {\bibinfo {author} {\bibfnamefont {S.}~\bibnamefont
  {Succi}}, \bibinfo {author} {\bibfnamefont {F.}~\bibnamefont
  {Fillion-Gourdeau}}, \ and\ \bibinfo {author} {\bibfnamefont
  {S.}~\bibnamefont {Palpacelli}},\ }\href {\doibase
  10.1140/epjqt/s40507-015-0025-1} {\bibfield  {journal} {\bibinfo  {journal}
  {EPJ Quantum Technology}\ }\textbf {\bibinfo {volume} {2}},\ \bibinfo {pages}
  {1} (\bibinfo {year} {2015})}\BibitemShut {NoStop}%
\bibitem [{\citenamefont {Weinheimer}(2010)}]{Weinheimer2010}%
  \BibitemOpen
  \bibfield  {author} {\bibinfo {author} {\bibfnamefont {C.}~\bibnamefont
  {Weinheimer}},\ }\href {\doibase
  http://dx.doi.org/10.1016/j.ppnp.2009.12.011} {\bibfield  {journal} {\bibinfo
   {journal} {Progress in Particle and Nuclear Physics}\ }\textbf {\bibinfo
  {volume} {64}},\ \bibinfo {pages} {205 } (\bibinfo {year} {2010})},\ \bibinfo
  {note} {neutrinos in Cosmology, in Astro, Particle and Nuclear
  PhysicsInternational Workshop on Nuclear Physics, 31st course}\BibitemShut
  {NoStop}%
\bibitem [{\citenamefont {Noh}\ \emph {et~al.}(2011)\citenamefont {Noh},
  \citenamefont {Rodr{\'i}guez-Lara},\ and\ \citenamefont
  {Angelakis}}]{Noh2011}%
  \BibitemOpen
  \bibfield  {author} {\bibinfo {author} {\bibfnamefont {C.}~\bibnamefont
  {Noh}}, \bibinfo {author} {\bibfnamefont {B.~M.}\ \bibnamefont
  {Rodr{\'i}guez-Lara}}, \ and\ \bibinfo {author} {\bibfnamefont {D.~G.}\
  \bibnamefont {Angelakis}},\ }\href@noop {} {\bibfield  {journal} {\bibinfo
  {journal} {New J. Phys.}\ }\textbf {\bibinfo {volume} {14,}},\ \bibinfo
  {pages} {033028} (\bibinfo {year} {2011})},\ \Eprint
  {http://arxiv.org/abs/1108.0182} {1108.0182} \BibitemShut {NoStop}%
\bibitem [{\citenamefont {Marini}\ \emph {et~al.}(2014)\citenamefont {Marini},
  \citenamefont {Longhi},\ and\ \citenamefont
  {Biancalana}}]{PhysRevLett.113.150401}%
  \BibitemOpen
  \bibfield  {author} {\bibinfo {author} {\bibfnamefont {A.}~\bibnamefont
  {Marini}}, \bibinfo {author} {\bibfnamefont {S.}~\bibnamefont {Longhi}}, \
  and\ \bibinfo {author} {\bibfnamefont {F.}~\bibnamefont {Biancalana}},\
  }\href {\doibase 10.1103/PhysRevLett.113.150401} {\bibfield  {journal}
  {\bibinfo  {journal} {Phys. Rev. Lett.}\ }\textbf {\bibinfo {volume} {113}},\
  \bibinfo {pages} {150401} (\bibinfo {year} {2014})}\BibitemShut {NoStop}%
\bibitem [{\citenamefont {Lan}\ \emph {et~al.}(2011)\citenamefont {Lan},
  \citenamefont {Celi}, \citenamefont {Lu}, \citenamefont {\"Ohberg},\ and\
  \citenamefont {Lewenstein}}]{PhysRevLett.107.253001}%
  \BibitemOpen
  \bibfield  {author} {\bibinfo {author} {\bibfnamefont {Z.}~\bibnamefont
  {Lan}}, \bibinfo {author} {\bibfnamefont {A.}~\bibnamefont {Celi}}, \bibinfo
  {author} {\bibfnamefont {W.}~\bibnamefont {Lu}}, \bibinfo {author}
  {\bibfnamefont {P.}~\bibnamefont {\"Ohberg}}, \ and\ \bibinfo {author}
  {\bibfnamefont {M.}~\bibnamefont {Lewenstein}},\ }\href {\doibase
  10.1103/PhysRevLett.107.253001} {\bibfield  {journal} {\bibinfo  {journal}
  {Phys. Rev. Lett.}\ }\textbf {\bibinfo {volume} {107}},\ \bibinfo {pages}
  {253001} (\bibinfo {year} {2011})}\BibitemShut {NoStop}%
\bibitem [{\citenamefont {Goyal}\ \emph {et~al.}(2015)\citenamefont {Goyal},
  \citenamefont {Roux}, \citenamefont {Forbes},\ and\ \citenamefont
  {Konrad}}]{PhysRevA.92.040302}%
  \BibitemOpen
  \bibfield  {author} {\bibinfo {author} {\bibfnamefont {S.~K.}\ \bibnamefont
  {Goyal}}, \bibinfo {author} {\bibfnamefont {F.~S.}\ \bibnamefont {Roux}},
  \bibinfo {author} {\bibfnamefont {A.}~\bibnamefont {Forbes}}, \ and\ \bibinfo
  {author} {\bibfnamefont {T.}~\bibnamefont {Konrad}},\ }\href {\doibase
  10.1103/PhysRevA.92.040302} {\bibfield  {journal} {\bibinfo  {journal} {Phys.
  Rev. A}\ }\textbf {\bibinfo {volume} {92}},\ \bibinfo {pages} {040302}
  (\bibinfo {year} {2015})}\BibitemShut {NoStop}%
\bibitem [{\citenamefont {Mallick}\ \emph {et~al.}(2016)\citenamefont
  {Mallick}, \citenamefont {Mandal},\ and\ \citenamefont
  {Chandrashekar}}]{Mallick2016}%
  \BibitemOpen
  \bibfield  {author} {\bibinfo {author} {\bibfnamefont {A.}~\bibnamefont
  {Mallick}}, \bibinfo {author} {\bibfnamefont {S.}~\bibnamefont {Mandal}}, \
  and\ \bibinfo {author} {\bibfnamefont {C.~M.}\ \bibnamefont
  {Chandrashekar}},\ }\href@noop {} {\  (\bibinfo {year} {2016})},\ \Eprint
  {http://arxiv.org/abs/1604.04233} {1604.04233} \BibitemShut {NoStop}%
\bibitem [{\citenamefont {Bernab{\'e}u}\ \emph {et~al.}(2002)\citenamefont
  {Bernab{\'e}u}, \citenamefont {Palomares-Ruiz}, \citenamefont {P{\'e}rez},\
  and\ \citenamefont {Petcov}}]{Bernabeu2002}%
  \BibitemOpen
  \bibfield  {author} {\bibinfo {author} {\bibfnamefont {J.}~\bibnamefont
  {Bernab{\'e}u}}, \bibinfo {author} {\bibfnamefont {S.}~\bibnamefont
  {Palomares-Ruiz}}, \bibinfo {author} {\bibfnamefont {A.}~\bibnamefont
  {P{\'e}rez}}, \ and\ \bibinfo {author} {\bibfnamefont {S.}~\bibnamefont
  {Petcov}},\ }\href
  {http://www.sciencedirect.com/science/article/pii/S0370269302013588}
  {\bibfield  {journal} {\bibinfo  {journal} {Physics Letters B}\ }\textbf
  {\bibinfo {volume} {531}},\ \bibinfo {pages} {90} (\bibinfo {year}
  {2002})}\BibitemShut {NoStop}%
\bibitem [{\citenamefont {Feldman}\ \emph {et~al.}(2013)\citenamefont
  {Feldman}, \citenamefont {Hartnell},\ and\ \citenamefont
  {Kobayashi}}]{Feldman:2013vca}%
  \BibitemOpen
  \bibfield  {author} {\bibinfo {author} {\bibfnamefont {G.~J.}\ \bibnamefont
  {Feldman}}, \bibinfo {author} {\bibfnamefont {J.}~\bibnamefont {Hartnell}}, \
  and\ \bibinfo {author} {\bibfnamefont {T.}~\bibnamefont {Kobayashi}},\
  }\href@noop {} {\bibfield  {journal} {\bibinfo  {journal} {Adv.High Energy
  Phys.}\ }\textbf {\bibinfo {volume} {2013}},\ \bibinfo {pages} {475749}
  (\bibinfo {year} {2013})},\ \Eprint {http://arxiv.org/abs/1210.1778}
  {arXiv:1210.1778 [hep-ex]} \BibitemShut {NoStop}%
\bibitem [{Note1()}]{Note1}%
  \BibitemOpen
  \bibinfo {note} {We do not include CP symmetry violation effects in this
  model.}\BibitemShut {Stop}%
\bibitem [{\citenamefont {Forero}\ \emph {et~al.}(2014)\citenamefont {Forero},
  \citenamefont {Tortola},\ and\ \citenamefont {Valle}}]{Forero2014}%
  \BibitemOpen
  \bibfield  {author} {\bibinfo {author} {\bibfnamefont {D.~V.}\ \bibnamefont
  {Forero}}, \bibinfo {author} {\bibfnamefont {M.}~\bibnamefont {Tortola}}, \
  and\ \bibinfo {author} {\bibfnamefont {J.~W.~F.}\ \bibnamefont {Valle}},\
  }\href@noop {} {\bibfield  {journal} {\bibinfo  {journal} {Phys. Rev. D}\
  }\textbf {\bibinfo {volume} {90,}},\ \bibinfo {pages} {093006} (\bibinfo
  {year} {2014})},\ \Eprint {http://arxiv.org/abs/1405.7540} {1405.7540}
  \BibitemShut {NoStop}%
\bibitem [{inv()}]{invisibleseu}%
  \BibitemOpen
  \href
  {http://invisibles.eu/outreach/entry/ceaseless-transformation-three-neutrinos}
  {\emph {\bibinfo {title} {Invisibles: Neutrino, dark matter and dark energy
  physics}}},\ \bibinfo {address} {http://invisibles.eu}\BibitemShut {NoStop}%
\bibitem [{\citenamefont {Wolfenstein}(1978)}]{Wolfenstein:1978ue-pot}%
  \BibitemOpen
  \bibfield  {author} {\bibinfo {author} {\bibfnamefont {L.}~\bibnamefont
  {Wolfenstein}},\ }\href@noop {} {\bibfield  {journal} {\bibinfo  {journal}
  {Phys. Rev.}\ }\textbf {\bibinfo {volume} {D17}},\ \bibinfo {pages} {2369}
  (\bibinfo {year} {1978})}\BibitemShut {NoStop}%
\bibitem [{\citenamefont {Mikheev}\ and\ \citenamefont
  {Smirnov}(1985)}]{Mikheev1985}%
  \BibitemOpen
  \bibfield  {author} {\bibinfo {author} {\bibfnamefont {S.}~\bibnamefont
  {Mikheev}}\ and\ \bibinfo {author} {\bibfnamefont {A.}~\bibnamefont
  {Smirnov}},\ }\bibfield  {booktitle} {\emph {\bibinfo {booktitle} {Soviet
  Journal of Nuclear Physics}},\ }\href@noop {} {\ \textbf {\bibinfo {volume}
  {42}},\ \bibinfo {pages} {913} (\bibinfo {year} {1985})}\BibitemShut
  {NoStop}%
\bibitem [{\citenamefont {Petcov}(1997)}]{Petcov1997}%
  \BibitemOpen
  \bibfield  {author} {\bibinfo {author} {\bibfnamefont {S.}~\bibnamefont
  {Petcov}},\ }\href {\doibase http://dx.doi.org/10.1016/S0370-2693(97)00704-1}
  {\bibfield  {journal} {\bibinfo  {journal} {Physics Letters B}\ }\textbf
  {\bibinfo {volume} {406}},\ \bibinfo {pages} {355 } (\bibinfo {year}
  {1997})}\BibitemShut {NoStop}%
\end{thebibliography}
\end{document}